\documentclass[aps,prl,reprint,superscriptaddress,showpacs,preprintnumbers,amsmath,amssymb]{revtex4-1}
\usepackage{graphicx}
\graphicspath{{fig/}}
\usepackage{dcolumn}
\usepackage{bm}
\usepackage{color}
\usepackage{amssymb}
\usepackage{amsmath}
\usepackage{soul}
\begin{document}

%
\title{Collective motion of macroscopic spheres floating on capillary ripples:\\
Dynamic heterogeneity and dynamic criticality}

\author{Ceyda Sanl{\i}}
\email{cedaysan@gmail.com}
\altaffiliation{Present address: CompleXity Networks, naXys, University of Namur, 5000 Namur, Belgium}
\affiliation{Physics of Fluids, University of Twente, P.O. Box 217, 7500 AE Enschede, The Netherlands}

\author{Kuniyasu Saitoh}
\email{k.saitoh@utwente.nl}
\affiliation{Multi Scale Mechanics, University of Twente, P.O. Box 217, 7500 AE Enschede, The Netherlands}

\author{Stefan Luding}
\email{S.Luding@utwente.nl} \affiliation{Multi Scale
Mechanics, University of Twente, P.O. Box 217, 7500 AE Enschede,
The Netherlands}

\author{Devaraj van der Meer}
\email{d.vandermeer@tnw.utwente.nl}
\affiliation{Physics of Fluids, University of Twente, P.O. Box 217, 7500 AE Enschede, The Netherlands}

\date{\today}

\begin{abstract}


When a dense monolayer of macroscopic spheres floats on
chaotic capillary Faraday waves, a coexistence of large scale
convective motion and caging dynamics typical for jammed systems
is observed.
We subtract the convective mean flow using a homogenization or coarse graining method and reveal subdiffusion for the caging time scales 
followed by a diffusive regime at later times. We apply the
methods of dynamic heterogeneity and show that the typical
time and length scales of the fluctuations due to rearrangements
of observed particle groups significantly increase when the system
approaches its densest 
experimentally accessible 
concentration. To connect 
the system to the 
dynamic criticality literature
we fit power laws to our results.
The resultant critical exponents are consistent with those
found in dense suspensions of colloids indicating universal
stochastic dynamics.

\end{abstract}







\maketitle
Small-scale events can dominate statistical systems 
to such an extent that one observes 
 phenomena on a global scale. From the classical to the quantum limit, 
microscopic fluctuations may even change the phase of matter when appropriate control parameters are tuned to critical values \cite{Pathria,fluctuation_Berthier_NaturePhys}. Even if their origin and nature is not always 
understood, these spatiotemporal microscopic fluctuations can drive common observable behavior near to such 
a phase transition. 
For classical particulate systems, a vast range of materials exhibits a sudden change to a rigid state called 
a glass or jamming transition. Thermal systems, e.g., supercooled liquids at a critical temperature, or emulsions and colloidal suspensions at a critical packing fraction \cite{glass0,glass1,glass2,glass3}, exhibit a glass transition. Furthermore, athermal systems such as foams and granulates experience a jamming transition, also at a critical packing fraction \cite{ph0,gn0,gn1,gn2,gn3}. In all these systems, transient spatial fluctuations lead to a 
large scale cooperative motion of their constituents 
near the transition \cite{glass0,NarumiFranklinDesmondTokuyamaWeeks,RahmaniVaartDamHuChikkadiSchall,shear0,air0,air2,horiz1,horiz0,foamco}.

In this Letter, we investigate the dynamics of the collective events near the jamming transition in an alternative experiment: Macroscopic spheres floating on the surface of capillary Faraday waves. Our control parameter is the floating sphere concentration $\phi$ on the surface 
which is varied from a moderate value to the maximum value attainable 
experimentally. Erratic forces due to the surface waves \cite{chaoticwaves} and the attractive capillary interaction among the spheres \cite{Cheerios,Capillaryforce} make our system markedly different from the 
previously studied ones 
\cite{2d_driven_GM_simulation,air0,shear0,shear2,horiz1}: A distinct feature is 
a large scale convection of the spheres on the wave which (for all $\phi$) forms naturally and strongly affects the visible dynamics. We aim to understand to what extent concepts from the glass and jamming literature --such as dynamical heterogeneity (DH) and dynamical criticality (DC)--  still hold in this convective system. To do so we subtract the convective mean flow using a coarse graining (CG) method and analyze the features of our system both before and after this procedure.

Dynamical heterogeneity (DH) investigates the relation between the local dynamic events on the particle scale 
and the resultant large-scale cooperative motion 
\cite{glass3,BerthierPRE2004,BerthierBiroliBouchaudCipellettiMasriLHoteLadieuPierno,glass1,BerthierBiroliBouchaudCipellettiSaarloos,Hatano_scaling_heterogeneity}. For its quantification 
two observables are 
introduced: The dynamic susceptibility, a measure of to what extent the dynamics of the system is heterogeneous in space and irregular in time, and the four-point correlation function, a measure of how often and from how far two arbitrarily chosen locally heterogenous events correlate to one another in time. 
Both quantities 
were calculated for colloids \cite{glass0,NarumiFranklinDesmondTokuyamaWeeks,RahmaniVaartDamHuChikkadiSchall}, driven hard granulates \cite{shear0,shear1,shear2,shear3,air0,air1,air2}, and foams \cite{foamco}. For all cases, the length scales, time scales, and the 
number of collective events, such as rearrangements of particle groups, 
dramatically increase near the transition.


The common nature of the behavior of classical particulate systems near 
transitions encourages to ask whether there is universality. 
This has led to the concept of 
dynamical criticality (DC) 
\cite{DC-refI,DC-refII,DC_refIII}. 
Briefly, DC postulates 
a power-law relation between the (diverging) length and time scales 
close to the phase transition. The uniqueness of this --and other-- exponents in different systems would then support the existence of universality \cite{DC-refI,DC-refII,DC_refIII}.
There is 
evidence pointing to universality in the above sense in various systems \cite{air0,air2,horiz1,shear2}. However, 
investigation is ongoing \cite{Hatano_scaling_growingscales,Brian_scaling_fluctuations,Hatano_criticalscaling,Hatano_scaling_heterogeneity} 
and increasing the 
number of systems either obeying or disobeying the universality is key to reaching a more complete understanding. We will therefore analyze our system in the light of both DH and DC. 


\begin{figure}
\includegraphics[width= 9cm]{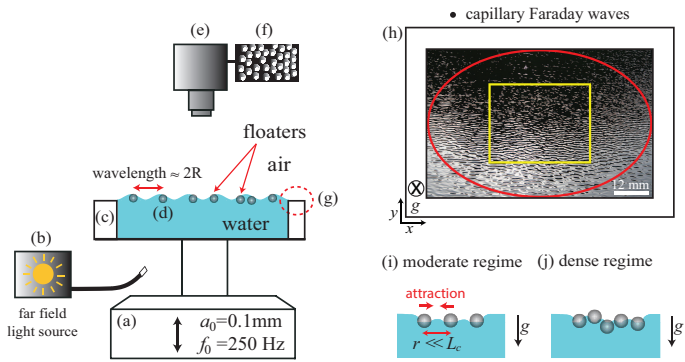}
\caption{(Color online)
Left panel: Experimental setup: (a) Shaker, 
(b) far field light source, 
(c) transparent hydrophilic glass container, (d) hydrophilic 
spheres, (e) high speed camera, 
(f) sample from 
a camera image for dense floater concentrations. The advantage of using far field light source is that 
all particles can be detected using the white spots in the images. (g) Pinned brim-full boundary condition. Right panel: (h) Top view of 
capillary 
Faraday waves, with the elliptic boundary containing the particles (red ellipse) and the field of view of the camera 
(yellow rectangle). (i-j) Sketch of the typical meniscus around the hydrophilic floaters in moderate (i) and extremely dense (j) regimes.\label{fig:setup}}
\end{figure}

A schematic illustration of the experimental setup is shown in Fig.\ \ref{fig:setup}. A rectangular container [Fig.\ \ref{fig:setup}(a)] is attached to a shaker providing a vertical sinusoidal oscillation such that the vertical position of the container varies as a function of time $t$ as $a_0\sin (2\pi f_0\,t)$, where $a_0$ is the shaking amplitude and $f_0$ is the shaking frequency. Here, both $a_0$ and $f_0$ are fixed to 0.1 mm and 250 Hz, respectively. This combination is chosen to create capillary ripples on the water surface with a wavelength in the order of the floater diameter ($\approx0.62$ mm). The container is 
filled with purified water (Millipore water with a resistivity $>$ 18 M$\Omega\cdot$cm) such that the water level is perfectly matched with the container edge as shown in Fig.\ \ref{fig:setup}(g) to create the brim-full boundary condition \cite{Douady}. Spherical hydrophilic polystyrene floaters [Fig.\ \ref{fig:setup}(d)], contact angle $74^\circ$ and density 1050 kg/m$^3$, with an average radius $R$ of 0.31 mm are carefully distributed over the water surface to make a monolayer of floaters \footnote{The particles have been custom made by a collaborating company and are not commercially available.}. The polydispersity of the floaters is approximately $14\%$ and assumed to be just wide enough to avoid crystallization \cite{poly_luding}. To avoid any surfactant effects, both the container and the floaters are cleaned by performing the cleaning protocol as described in Ref. \cite{cleaningprotocol}.

A continuous white fiber light source (Schott) is used to illuminate the floaters from far away as shown in Fig.\ \ref{fig:setup}(b). The positions of the floaters are recorded with a high-speed camera (Photron Fastcam SA.1) at 60-500 frames per second. The lens (Carl Zeiss 60mm) is adjusted such that it focuses on the floaters at the non-deformed water surface. Here, we use the random capillary Faraday waves to just agitate the dense floaters so that there is no macroscopic apparent amplitude observed. The wave amplitude is always considerably smaller than the floater radius ($\approx0.31$ mm).

The resultant capillary ripples on the water surface in the container, made from transparent hydrophilic glass with 10 mm height and a 81$\times$45 mm$^2$ rectangular cross section, are shown in Fig.\ \ref{fig:setup}(h). To eliminate the boundary effects due to the sharp corners of the container, an elliptic rim made from plastic 
is used to contain the particles. Each image taken with the high speed camera is 512 $\times$ 640 px$^2$ (36$\times$28 mm$^2$), where px means pixel, as shown by the yellow rectangle (size ratios are preserved). The horizontal field of view is $\sim 35\%$ of the total area of the ellipse. 
Due to the asymmetric surface deformation around each hydrophilic
heavy sphere, there is an attraction \cite{Cheerios,
Capillaryforce} between the spheres so that the floaters are
cohesive \footnote{Even in the dilute case, the distance $r$
between the floaters is much smaller than the capillary length
$L_c=\sqrt{\gamma/\rho_l g}$ with $\gamma$ the surface tension
coefficient of the interface, $\rho_l$ the liquid density, and $g$
the acceleration of gravity. For an air-water interface at
$20\,^\circ$C, $L_c=2.7$ mm.}. For moderate $\phi$, the monolayer
can be considered two-dimensional [Fig.~
\ref{fig:setup}(i)]. In the dense regime however [Fig.~\ref{fig:setup}(j)], particles are so close
that the layer may (locally) buckle and have
three-dimensional aspects \cite{bucklingII}.

The control parameter of the experiment is the floater concentration $\phi$, which (ignoring buckling) is measured by determining the area fraction covered by the floaters in the area of interest [Fig.\ \ref{fig:setup}(h)]. 
In this study, $\phi$ is increased from moderate to dense concentrations, 
$\phi=0.65-0.77$. 

Under the influence of the erratic capillary waves and the attractive capillary interaction,
a large scale convective motion is observed with a typical length scale 
$\sim 60$ times larger
than the floater diameter, which is $\sim1/2$ of the system size, and a typical time scale 
$\sim 250$ times longer than that of
the capillary Faraday waves. To focus on microscopic fluctuations, we subtract the displacement due to 
the large scale convective motion from our experimental data.
At first, we define the velocity field by the coarse graining (CG) method \cite{coarse6,coarse9,coarse10} as
\begin{equation}
\mathbf{u}(\mathbf{x},t)=\frac{\sum_i\mathbf{v}_i(t)
\psi_d(|\mathbf{x}-\mathbf{x}_i(t)|)}{\sum_i\psi_d(|\mathbf{x}-\mathbf{x}_i(t)|)}
\label{eq:vfield}
\end{equation}
with the position $\mathbf{x}_i(t)$ and the velocity $\mathbf{v}_i(t)$ of the $i$-th floater,
where we adopt both Gaussian, $e^{-(x/d)^2}$,
and Heaviside, $\Theta_d(x)=1$ ($x\le d$) and zero otherwise, 
as CG kernel functions $\psi_d(x)$. 
Here, $d$ is a length scale of the order of the particle diameter. Subsequently, 
we subtract the displacement $\mathbf{l}_i(t)=\int_0^t\mathbf{u}(\mathbf{x}_i(s),s)ds$ due to this macroscopic flow from the position as
$\mathbf{r}_i(t)\equiv\mathbf{x}_i(t)-\mathbf{l}_i(t)$ and define an actual displacement during the time interval $\tau$ as
$D_i(t,\tau)\equiv|\mathbf{r}_i(t+\tau)-\mathbf{r}_i(t)|$.

First, we look at single particle dynamics. Approaching the maximal density, particles experience 
a cage effect, i.e., they are locally trapped by the nearest neighbors, a cage from which, 
in the presence of fluctuations, sudden escape 
jumps 
may occur \cite{glass0}. The caging and the jumps leave their tracks in 
the mean square displacement of individual particles as a subdiffusion regime for short times and ordinary diffusion at later times.
For our system, we find very similar behavior.

\begin{figure}[ht!]
\includegraphics[width= 0.5\textwidth]{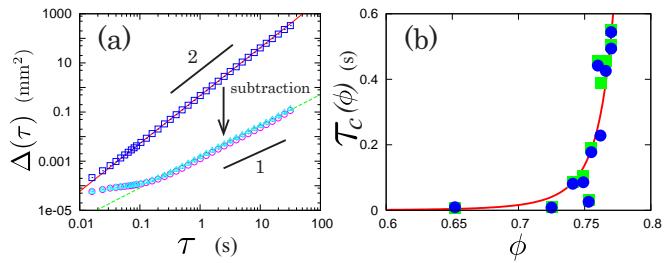}
\caption{(Color online)
(a) Mean square displacement (MSD) of the floaters for $\phi=0.755$. The open squares are
obtained without subtracting the displacement $\mathbf{l}_i(t)$ due to the large scale convection. 
The open circles and open triangles are the results with the subtraction, where we use a Gaussian with $d=\sigma$ and a step function with $d=1.6\sigma$ as coarse graining (CG) functions, respectively. Here, $\sigma$ is the floater diameter.
The red solid and the green dotted lines have slopes $1.9$ and $1.1$, respectively.
(b) Crossover time $\tau_c$ plotted against $\phi$, where the closed circles and 
squares are obtained using the 
Gaussian ($d=\sigma$) and the step 
($d=1.6\sigma$)
CG functions, respectively. 
The solid line represents $\tau_c\sim(\phi_J-\phi)^{-3.9}$ with $\phi_J=0.82$.
\label{fig:msd}}
\end{figure}

Fig.\ \ref{fig:msd}(a) shows the mean square displacement (MSD) of the floaters
$\Delta(\tau)=\left\langle\sum_iD_i^2(t,\tau)/N\right\rangle_t$, where the brackets $\langle\dots\rangle_t$
represent an average over time $t$ \footnote{The procedure in the ensemble average is to calculate
$|\mathbf{r}_i(t+\tau)-\mathbf{r}_i(t)|$ using arbitrary starting times $t$ and averaging over t.
A similar procedure is followed in calculating ensemble averages in the self-overlap order parameter,
dynamic susceptibility and four-point correlation function.} and $N$ is the number of floaters in the sample.
In our experiment, the floaters are transported by the large scale convection, and thus, the resultant motion
is always ballistic. Therefore, when we do not subtract the displacement $\mathbf{l}_i(t)$ from the experimental
data, the MSD quadratically increases with time with a slope $2$ in the log-log plot [open squares in Fig.\ \ref{fig:msd}(a)].
However, when we do subtract the additional
displacement due to the convection for a suitable value of $d$ \footnote{The optimal values for $d$
were obtained as follows: When looking at the subtraction procedure as a function of $d$ we find that
the displacement rises steeply from zero for $d\ll\sigma$, $\sigma$ is the floater diameter,
into a plateau from which it continues to rise. A value in the center of the plateau is chosen,
which happens to correspond roughly to the particle diameter.}, both the initial subdiffusive and the later diffusive regimes are found.

As shown in Fig.\ \ref{fig:msd}(b), the crossover time $\tau_c$ 
between these subdiffusive and diffusive regimes, 
rapidly increases with $\phi$. Since the subdiffusion represents the cage effect of the floaters, it is plausible that 
the crossover time diverges at 
the jamming point, where no floater can ever escape from the cage.


On physical grounds, the jamming point $\phi_J$ needs to be smaller than the random close packing of slightly polydisperse disks, i.e., $\phi_J<0.84$ \footnote{From the (two-dimensional) pair correlation function $g(r)$ we observe no evidence for significant crystallization which may cause an increase of this upper limit for $\phi_J$. Secondly, although buckling may be a significant factor, it does not lead to a broadening of the first peak in $g(r)$ that one would expect to be present if particles start to overlap for increasing $\phi$. And, finally, the homogenized local packing fraction shows a sharp cut-off at $\phi\approx0.84$. These facts together suggests that $\phi_J<0.84$.}. In addition, it needs to be considerably larger than $\phi=0.77$, the largest experimental average at which we were able to measure. By fitting a power law $\tau_c\sim(\phi_J-\phi)^{\alpha}$ to our data \footnote{Due to their limited range other functional forms could possibly also fit our dynamic time and length scales. However, we restrict ourselves to power-law fits in order to compare our results to those in the literature.} we find that $\phi_J \approx 0.82$, which is consistent with the above and leads us to conclude that $\phi_J = 0.82\pm0.02$. Note that this value is considerably larger than the suggested static buckling density of the attractive monodisperse spheres~\cite{bucklingI}, namely $\phi_b\simeq0.71$. Next, we use this fixed value for $\phi_J$ in our power-law fit to obtain the exponent $\alpha\simeq-3.9\pm0.9$, which is consistent with the exponent $\alpha \simeq -4.0\pm0.6$ found
in an earlier experiment~\cite{shear2}.   

\begin{figure*}[ht!]
\includegraphics[width= 1.0 \textwidth]{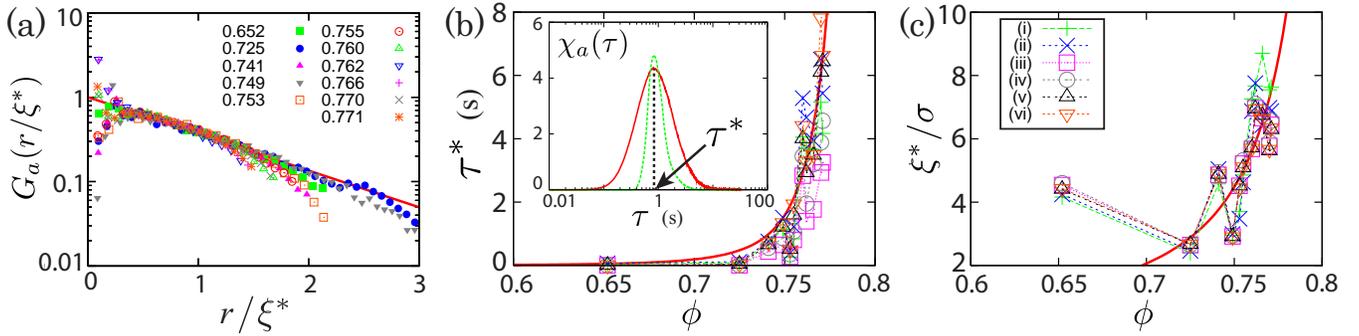}
\caption{(Color online)
(a) Scaled four-point correlation functions $G_a(r/\xi^\ast)\equiv (r/\xi^\ast)^{\beta}g_a(r,\tau^\ast)/A$ where we use Gaussians for the coarse graining and overlap functions respectively [case (iii) in Table \ref{tab:CGOF}] and the solid line represents $e^{-r/\xi^\ast}$.
(b) The time scale of the dynamic heterogeneity $\tau^\ast$ vs. $\phi$ obtained from the dynamic susceptibility $\chi_a(\tau)$. The solid line represents $\tau^\ast\sim(\phi_J-\phi)^{-3.9}$. 
Inset: Dynamic susceptibilities, where the green dotted line is obtained without subtracting the convective displacement $\mathbf{l}_i(t)$ [case (i) in Table \ref{tab:CGOF}] and the red solid line with the subtraction [case (iii) in Table \ref{tab:CGOF}]. (c) The dynamic correlation length $\xi^\ast$ obtained from $g_a(r,\tau)$ for $\tau=\tau^\ast$. The solid line represents
$\xi^\ast\sim(\phi_J-\phi)^{-1.4}$. 
The legend in (c) indicates the conditions described in Table \ref{tab:CGOF} and explains the symbols in (b, c). \label{fig:g4_tau_xsi}}
\end{figure*}

To quantify the heterogenous dynamics of the floaters, we introduce the self-overlap order parameter $q_a(t,\tau)=\sum_iw_a(D_i(t,\tau))/N$ and the dynamic susceptibility $\chi_a(\tau)=N\left[\langle q^2_a(t,\tau)\rangle_t-\langle q_a(t,\tau)\rangle_t^2\right]$. Here, $w_a(x)$ is the overlap function (OF) defined as 
a Gaussian 
$e^{-(x/a)^2}$ or 
a Heaviside step function 
$\Theta_a(x)$ as defined previously ($1$ for $x\le a$ and $0$ otherwise).
The width of the OF, $a$, is a measure for the typical distance over which a single floater can move within time $\tau$. To disregard the motion of the floaters in the cage, $a$ is chosen to be larger than 
their typical displacement inside a cage and also chosen to maximize the extremal value of the dynamic susceptibility as shown in the inset of Fig.\ \ref{fig:g4_tau_xsi}(b) 
\cite{horiz1}.


The various coarse graining (CG) functions and overlap functions (OF) in total give us six different manners of analyzing the data, if we also include the possibility of \emph{not} subtracting the displacement due to the large scale convection before the DH analysis. These are summarized in Table \ref{tab:CGOF}, together with the optimal values of $d$ and $a$ obtained as described above \footnote{Both $d$ and $a$ are first calculated for each $\phi$ separately, and then appropriate values are obtained by averaging over the determined values.}. When we plot the dynamic susceptibility $\chi_a(\tau)$ we obtain 
similar results in all six cases [the inset of Fig.\ \ref{fig:g4_tau_xsi}(b)]. In particular the location of the peak in $\chi_a(\tau)$ provides us with an estimate of the typical time scale $\tau^\ast$ of the dynamic heterogeneity, which are plotted for all six cases as functions of $\phi$ in Fig.\ \ref{fig:g4_tau_xsi}(b).


\begin{table}
\caption{Analysis methods: The set of the CG function and the OF, with 
$d$ and $a$ 
as described in the main text, both in terms of 
the floater diameter $\sigma$.
\label{tab:CGOF}}
\begin{ruledtabular}
\begin{tabular}{ccccc}
& CG function &  OF & $d/\sigma$ & $a/\sigma$\\
\hline
(i)  & None      & Gaussian  & $-$ & $0.49$ \\
(ii) & None      & Heaviside & $-$ & $0.52$ \\
(iii)& Gaussian  & Gaussian  & $1.0$ & $0.038$ \\
(iv) & Gaussian  & Heaviside & $1.0$ & $0.042$ \\
(v)  & Heaviside & Gaussian  & $1.6$ & $0.042$ \\
(vi) & Heaviside & Heaviside & $1.6$ & $0.046$ \\
\end{tabular}
\end{ruledtabular}
\end{table}

To investigate the dynamic correlation length of the floaters,
we apply the four-point correlation function \cite{tech0}
\begin{equation}
g_a(r,\tau)=\frac{1}{2\pi rN}\left\langle\sum_{i,j}\delta(r-r_{ij}(t))
c_{ij}(t,\tau)\right\rangle_t-\rho\langle q_a(t,\tau) \rangle_t^2
\label{eq:g4}
\end{equation}
satisfying $\chi_a(\tau)=2\pi\int rg_a(r,\tau)dr$, where $\rho\equiv N/S$ and $S$ are the number density of the floaters and the area of interest, respectively. $N$ is the number of floaters as introduced previously. In addition, we define $r_{ij}(t)\equiv|\mathbf{r}_i(t)-\mathbf{r}_j(t)|$ and $c_{ij}(t,\tau)\equiv w_a(D_i(t,\tau))w_a(D_j(t,\tau))$. 
Furthermore, we assume the Ornstein-Zernike form of the four-point correlation function \cite{tech0}, in which the dynamic correlation length $\xi^\ast$ is obtained considering the scaling $g_a(r,\tau^\ast)=A (r/\xi^\ast)^{-\beta}e^{-r/\xi^\ast}$ for some 
amplitude $A$ and exponent $\beta$, where $\tau^\ast$ is the time scale obtained from $\chi_a(\tau)$. 

Fig.\ \ref{fig:g4_tau_xsi}(a) shows the function $G_a(r/\xi^\ast)\equiv (r/\xi^\ast)^{\beta}g_a(r,\tau^\ast)/A$,
where the (very weak) exponent $\beta=0.01$ is taken to be independent of $\phi$. 
The resultant $G_a(r/\xi^\ast)$ 
successfully collapses onto a single master curve $e^{-r/\xi^\ast}$ for each $\phi$ except for the tails. 
This 
procedure is repeated for each condition 
in Table \ref{tab:CGOF}. Remarkably, we find that neither the value of the exponent nor the master curve presents any significant difference.

Fig.\ \ref{fig:g4_tau_xsi} displays the time scales of the dynamic heterogeneity, $\tau^\ast$ [Fig.\ \ref{fig:g4_tau_xsi}(b)], and the dynamic correlation
length, $\xi^\ast$ [Fig.\ \ref{fig:g4_tau_xsi}(c)], plotted versus $\phi$, where both increase strongly with $\phi$. One can introduce the power law fits 
\cite{shear2,DC-refI,DC-refII,DC_refIII}
\begin{eqnarray}
\tau^\ast=C(\phi_J-\phi)^{\eta}~,\label{eq:ta-phi}\\
\xi^\ast=D(\phi_J-\phi)^{\lambda}~\label{eq:xi-phi}.
\end{eqnarray}
Both the time exponent $\eta$ and the length exponent $\lambda$
are calculated considering all conditions reported in Table
\ref{tab:CGOF}. Fitting to the data we obtain $\eta\simeq-3.9\pm0.4$ and $\lambda\simeq-1.4\pm0.4$
for each condition in Table \ref{tab:CGOF} where we again used $\phi_J\simeq0.82\pm0.02$. 
Finally, combining Eqs. (\ref{eq:ta-phi}) and
(\ref{eq:xi-phi}) in the light of dynamical criticality
\cite{DC-refI,DC-refII,DC_refIII}, namely
$\tau^\ast\sim{\xi^\ast}^\nu$, we quantify the relation between
$\eta$ and $\lambda$ as $\nu=\eta/\lambda\simeq 2.7\pm1.2$.

Summarizing, by eliminating the naturally occurring large-scale convection of the floaters, 
we find from the mean square displacement that the single floater dynamics resembles the caging observed in glassy liquids, first with subdiffusion followed by normal diffusion for later times.
The crossover time $\tau_c(\phi)$ between the two regimes [Fig.\ \ref{fig:msd}(b)] 
diverges near the estimated jamming point $\phi_J$, which can
be fitted by a power law $(\phi_J-\phi)^\alpha$ with exponent
$\alpha\simeq-3.9$. A second time scale is that of the dynamic
heterogeneity $\tau^\ast$ [Fig.\ \ref{fig:g4_tau_xsi}(b)], which
can also be fitted by a power law which, remarkably but
consistently, has the very same exponent $\eta\simeq-3.9$.
The typical distance 
between 
two correlated, successive 
events, the dynamic
correlation length $\xi^\ast$ [Fig.\ \ref{fig:g4_tau_xsi}(c)] obtained from the four-point correlation function [Fig.\ \ref{fig:g4_tau_xsi}(a)], presents a power law scaling with exponent $\lambda\simeq-1.4$. 
Both 
of our dynamical exponents, $\eta$ and $\lambda$, 
are in a good agreement with the previous experiments on 
sheared microgel
spheres by Nordstrom \emph{et al}. \cite{shear2}, where the critical exponents of the time and length scales were found to be $-4$ and $-4/3$, respectively. 

The coarse graining procedure allows us to successfully remove the convective flow. While this is indeed necessary to study micro-fluctuation driven by diffusion, paradoxically, one of our main results is that it is not necessary to remove this mean flow for the dynamic susceptibility and the four-point correlation function. These results only depend insignificantly on whether the mean flow is subtracted or not. In addition, results do hardly depend on the choice of coarse graining and overlap functions, as long as the length scales in these test functions are optimized. In fact, from Table \ref{tab:CGOF} it can be appreciated that $a$ must be an order of magnitude lower with mean flow subtraction ($a\simeq0.04\sigma$) than without ($a\simeq0.5\sigma$). Finally, 
we determine from fits that $\phi_J=0.82$, which is considerably larger than the suggested critical density for static monodisperse 
floaters ($\phi_b=0.71$), and also than the largest 
concentration that we could reach experimentally, namely $\phi_{max}\simeq0.77$. For larger $\phi$, 
our layer of floating spheres is 
not stable under driving. Understanding the difference between $\phi_{max}$ and $\phi_J$  
requires further study and lies beyond the scope of this paper. 

\begin{acknowledgments}
We thank O. Dauchot, D. J. Durian, T. Hatano, H. Hayakawa, H. Katsuragi, T. Kawasaki, D. Lohse, K. Miyazaki, Y. Tagawa, and R. Yamamoto for fruitful discussions.
This work is partially supported by the NWO-STW VICI grant 10828 and is part of the research program of the FOM, which is financially supported by NWO; K. Saitoh and C. Sanl{\i} acknowledge financial support. C. Sanl{\i} acknowledges support from the Okinawa Inst. of Sci. and Tech. Graduate University (OIST).
\end{acknowledgments}
\bibliography{floats}

\begin{thebibliography}{53}%
\makeatletter
\providecommand \@ifxundefined [1]{%
 \@ifx{#1\undefined}
}%
\providecommand \@ifnum [1]{%
 \ifnum #1\expandafter \@firstoftwo
 \else \expandafter \@secondoftwo
 \fi
}%
\providecommand \@ifx [1]{%
 \ifx #1\expandafter \@firstoftwo
 \else \expandafter \@secondoftwo
 \fi
}%
\providecommand \natexlab [1]{#1}%
\providecommand \enquote  [1]{``#1''}%
\providecommand \bibnamefont  [1]{#1}%
\providecommand \bibfnamefont [1]{#1}%
\providecommand \citenamefont [1]{#1}%
\providecommand \href@noop [0]{\@secondoftwo}%
\providecommand \href [0]{\begingroup \@sanitize@url \@href}%
\providecommand \@href[1]{\@@startlink{#1}\@@href}%
\providecommand \@@href[1]{\endgroup#1\@@endlink}%
\providecommand \@sanitize@url [0]{\catcode `\\12\catcode `\$12\catcode
  `\&12\catcode `\#12\catcode `\^12\catcode `\_12\catcode `\%12\relax}%
\providecommand \@@startlink[1]{}%
\providecommand \@@endlink[0]{}%
\providecommand \url  [0]{\begingroup\@sanitize@url \@url }%
\providecommand \@url [1]{\endgroup\@href {#1}{\urlprefix }}%
\providecommand \urlprefix  [0]{URL }%
\providecommand \Eprint [0]{\href }%
\providecommand \doibase [0]{http://dx.doi.org/}%
\providecommand \selectlanguage [0]{\@gobble}%
\providecommand \bibinfo  [0]{\@secondoftwo}%
\providecommand \bibfield  [0]{\@secondoftwo}%
\providecommand \translation [1]{[#1]}%
\providecommand \BibitemOpen [0]{}%
\providecommand \bibitemStop [0]{}%
\providecommand \bibitemNoStop [0]{.\EOS\space}%
\providecommand \EOS [0]{\spacefactor3000\relax}%
\providecommand \BibitemShut  [1]{\csname bibitem#1\endcsname}%
\let\auto@bib@innerbib\@empty
\bibitem [{\citenamefont {Pathria}(1996)}]{Pathria}%
  \BibitemOpen
  \bibfield  {author} {\bibinfo {author} {\bibfnamefont {R.~K.}\ \bibnamefont
  {Pathria}},\ }\href@noop {} {\emph {\bibinfo {title} {Statistical
  Mechanics}}}\ (\bibinfo  {publisher} {Butterworth-Heinemann, Oxford $\&$
  Woburn},\ \bibinfo {year} {1996})\BibitemShut {NoStop}%
\bibitem [{\citenamefont {Berthier}\ and\ \citenamefont
  {Kurchan}(2013)}]{fluctuation_Berthier_NaturePhys}%
  \BibitemOpen
  \bibfield  {author} {\bibinfo {author} {\bibfnamefont {L.}~\bibnamefont
  {Berthier}}\ and\ \bibinfo {author} {\bibfnamefont {J.}~\bibnamefont
  {Kurchan}},\ }\href@noop {} {\bibfield  {journal} {\bibinfo  {journal}
  {Nature Phys.}\ }\textbf {\bibinfo {volume} {9}},\ \bibinfo {pages} {310}
  (\bibinfo {year} {2013})}\BibitemShut {NoStop}%
\bibitem [{\citenamefont {Berthier}\ and\ \citenamefont
  {Biroli}(2011)}]{glass0}%
  \BibitemOpen
  \bibfield  {author} {\bibinfo {author} {\bibfnamefont {L.}~\bibnamefont
  {Berthier}}\ and\ \bibinfo {author} {\bibfnamefont {G.}~\bibnamefont
  {Biroli}},\ }\href@noop {} {\bibfield  {journal} {\bibinfo  {journal} {Rev.\
  Mod.\ Phys.}\ }\textbf {\bibinfo {volume} {83}},\ \bibinfo {pages} {587}
  (\bibinfo {year} {2011})}\BibitemShut {NoStop}%
\bibitem [{\citenamefont {Berthier}\ \emph {et~al.}(2007)\citenamefont
  {Berthier}, \citenamefont {Biroli}, \citenamefont {Bouchaud}, \citenamefont
  {Kob}, \citenamefont {Miyazaki},\ and\ \citenamefont {Reichman}}]{glass1}%
  \BibitemOpen
  \bibfield  {author} {\bibinfo {author} {\bibfnamefont {L.}~\bibnamefont
  {Berthier}}, \bibinfo {author} {\bibfnamefont {G.}~\bibnamefont {Biroli}},
  \bibinfo {author} {\bibfnamefont {J.-P.}\ \bibnamefont {Bouchaud}}, \bibinfo
  {author} {\bibfnamefont {W.}~\bibnamefont {Kob}}, \bibinfo {author}
  {\bibfnamefont {K.}~\bibnamefont {Miyazaki}}, \ and\ \bibinfo {author}
  {\bibfnamefont {D.~R.}\ \bibnamefont {Reichman}},\ }\href@noop {} {\bibfield
  {journal} {\bibinfo  {journal} {J.\ Chem.\ Phys.}\ }\textbf {\bibinfo
  {volume} {126}},\ \bibinfo {pages} {184503} (\bibinfo {year}
  {2007})}\BibitemShut {NoStop}%
\bibitem [{\citenamefont {Ediger}(2000)}]{glass2}%
  \BibitemOpen
  \bibfield  {author} {\bibinfo {author} {\bibfnamefont {M.~D.}\ \bibnamefont
  {Ediger}},\ }\href@noop {} {\bibfield  {journal} {\bibinfo  {journal} {Ann.\
  Rev.\ Phys.\ Chem.}\ }\textbf {\bibinfo {volume} {51}},\ \bibinfo {pages}
  {99} (\bibinfo {year} {2000})}\BibitemShut {NoStop}%
\bibitem [{\citenamefont {Berthier}(2011)}]{glass3}%
  \BibitemOpen
  \bibfield  {author} {\bibinfo {author} {\bibfnamefont {L.}~\bibnamefont
  {Berthier}},\ }\href@noop {} {\bibfield  {journal} {\bibinfo  {journal}
  {Physics}\ }\textbf {\bibinfo {volume} {4}},\ \bibinfo {pages} {42} (\bibinfo
  {year} {2011})}\BibitemShut {NoStop}%
\bibitem [{\citenamefont {Liu}\ and\ \citenamefont {Nagel}(1998)}]{ph0}%
  \BibitemOpen
  \bibfield  {author} {\bibinfo {author} {\bibfnamefont {A.~J.}\ \bibnamefont
  {Liu}}\ and\ \bibinfo {author} {\bibfnamefont {S.~R.}\ \bibnamefont
  {Nagel}},\ }\href@noop {} {\bibfield  {journal} {\bibinfo  {journal} {Nature
  (London)}\ }\textbf {\bibinfo {volume} {396}},\ \bibinfo {pages} {21}
  (\bibinfo {year} {1998})}\BibitemShut {NoStop}%
\bibitem [{\citenamefont {O'Hern}\ \emph {et~al.}(2002)\citenamefont {O'Hern},
  \citenamefont {Langer}, \citenamefont {Liu},\ and\ \citenamefont
  {Nagel}}]{gn0}%
  \BibitemOpen
  \bibfield  {author} {\bibinfo {author} {\bibfnamefont {C.~S.}\ \bibnamefont
  {O'Hern}}, \bibinfo {author} {\bibfnamefont {S.~A.}\ \bibnamefont {Langer}},
  \bibinfo {author} {\bibfnamefont {A.~J.}\ \bibnamefont {Liu}}, \ and\
  \bibinfo {author} {\bibfnamefont {S.~R.}\ \bibnamefont {Nagel}},\ }\href@noop
  {} {\bibfield  {journal} {\bibinfo  {journal} {Phys.\ Rev.\ Lett.}\ }\textbf
  {\bibinfo {volume} {88}},\ \bibinfo {pages} {075507} (\bibinfo {year}
  {2002})}\BibitemShut {NoStop}%
\bibitem [{\citenamefont {O'Hern}\ \emph {et~al.}(2003)\citenamefont {O'Hern},
  \citenamefont {Silbert}, \citenamefont {Liu},\ and\ \citenamefont
  {Nagel}}]{gn1}%
  \BibitemOpen
  \bibfield  {author} {\bibinfo {author} {\bibfnamefont {C.~S.}\ \bibnamefont
  {O'Hern}}, \bibinfo {author} {\bibfnamefont {L.~E.}\ \bibnamefont {Silbert}},
  \bibinfo {author} {\bibfnamefont {A.~J.}\ \bibnamefont {Liu}}, \ and\
  \bibinfo {author} {\bibfnamefont {S.~R.}\ \bibnamefont {Nagel}},\ }\href@noop
  {} {\bibfield  {journal} {\bibinfo  {journal} {Phys.\ Rev.\ E}\ }\textbf
  {\bibinfo {volume} {68}},\ \bibinfo {pages} {011306} (\bibinfo {year}
  {2003})}\BibitemShut {NoStop}%
\bibitem [{\citenamefont {Majmudar}\ \emph {et~al.}(2007)\citenamefont
  {Majmudar}, \citenamefont {Sperl}, \citenamefont {Luding},\ and\
  \citenamefont {Behringer}}]{gn2}%
  \BibitemOpen
  \bibfield  {author} {\bibinfo {author} {\bibfnamefont {T.~S.}\ \bibnamefont
  {Majmudar}}, \bibinfo {author} {\bibfnamefont {M.}~\bibnamefont {Sperl}},
  \bibinfo {author} {\bibfnamefont {S.}~\bibnamefont {Luding}}, \ and\ \bibinfo
  {author} {\bibfnamefont {R.~P.}\ \bibnamefont {Behringer}},\ }\href@noop {}
  {\bibfield  {journal} {\bibinfo  {journal} {Phys.\ Rev.\ Lett.}\ }\textbf
  {\bibinfo {volume} {98}},\ \bibinfo {pages} {058001} (\bibinfo {year}
  {2007})}\BibitemShut {NoStop}%
\bibitem [{\citenamefont {van Hecke}(2010)}]{gn3}%
  \BibitemOpen
  \bibfield  {author} {\bibinfo {author} {\bibfnamefont {M.}~\bibnamefont {van
  Hecke}},\ }\href@noop {} {\bibfield  {journal} {\bibinfo  {journal} {J.\
  Phys.\ Cond.\ Mat.}\ }\textbf {\bibinfo {volume} {22}},\ \bibinfo {pages}
  {033101} (\bibinfo {year} {2010})}\BibitemShut {NoStop}%
\bibitem [{\citenamefont {Narumi}\ \emph {et~al.}(2011)\citenamefont {Narumi},
  \citenamefont {Franklin}, \citenamefont {Desmond}, \citenamefont {Tokuyama},\
  and\ \citenamefont {Weeks}}]{NarumiFranklinDesmondTokuyamaWeeks}%
  \BibitemOpen
  \bibfield  {author} {\bibinfo {author} {\bibfnamefont {T.}~\bibnamefont
  {Narumi}}, \bibinfo {author} {\bibfnamefont {S.~V.}\ \bibnamefont
  {Franklin}}, \bibinfo {author} {\bibfnamefont {K.~W.}\ \bibnamefont
  {Desmond}}, \bibinfo {author} {\bibfnamefont {M.}~\bibnamefont {Tokuyama}}, \
  and\ \bibinfo {author} {\bibfnamefont {E.~R.}\ \bibnamefont {Weeks}},\
  }\href@noop {} {\bibfield  {journal} {\bibinfo  {journal} {Soft Matter}\
  }\textbf {\bibinfo {volume} {7}},\ \bibinfo {pages} {1472} (\bibinfo {year}
  {2011})}\BibitemShut {NoStop}%
\bibitem [{\citenamefont {Rahmani}\ \emph {et~al.}(2012)\citenamefont
  {Rahmani}, \citenamefont {van~der Vaart}, \citenamefont {van Dam},
  \citenamefont {Hu}, \citenamefont {Chikkadi},\ and\ \citenamefont
  {Schall}}]{RahmaniVaartDamHuChikkadiSchall}%
  \BibitemOpen
  \bibfield  {author} {\bibinfo {author} {\bibfnamefont {Y.}~\bibnamefont
  {Rahmani}}, \bibinfo {author} {\bibfnamefont {K.}~\bibnamefont {van~der
  Vaart}}, \bibinfo {author} {\bibfnamefont {B.}~\bibnamefont {van Dam}},
  \bibinfo {author} {\bibfnamefont {Z.}~\bibnamefont {Hu}}, \bibinfo {author}
  {\bibfnamefont {V.}~\bibnamefont {Chikkadi}}, \ and\ \bibinfo {author}
  {\bibfnamefont {P.}~\bibnamefont {Schall}},\ }\href@noop {} {\bibfield
  {journal} {\bibinfo  {journal} {Soft Matter}\ }\textbf {\bibinfo {volume}
  {8}},\ \bibinfo {pages} {4264} (\bibinfo {year} {2012})}\BibitemShut
  {NoStop}%
\bibitem [{\citenamefont {Dauchot}\ \emph {et~al.}(2005)\citenamefont
  {Dauchot}, \citenamefont {Marty},\ and\ \citenamefont {Biroli}}]{shear0}%
  \BibitemOpen
  \bibfield  {author} {\bibinfo {author} {\bibfnamefont {O.}~\bibnamefont
  {Dauchot}}, \bibinfo {author} {\bibfnamefont {G.}~\bibnamefont {Marty}}, \
  and\ \bibinfo {author} {\bibfnamefont {G.}~\bibnamefont {Biroli}},\
  }\href@noop {} {\bibfield  {journal} {\bibinfo  {journal} {Phys.\ Rev.\
  Lett.}\ }\textbf {\bibinfo {volume} {95}},\ \bibinfo {pages} {265701}
  (\bibinfo {year} {2005})}\BibitemShut {NoStop}%
\bibitem [{\citenamefont {Keys}\ \emph {et~al.}(2007)\citenamefont {Keys},
  \citenamefont {Abate}, \citenamefont {Glotzer},\ and\ \citenamefont
  {Durian}}]{air0}%
  \BibitemOpen
  \bibfield  {author} {\bibinfo {author} {\bibfnamefont {A.~S.}\ \bibnamefont
  {Keys}}, \bibinfo {author} {\bibfnamefont {A.~R.}\ \bibnamefont {Abate}},
  \bibinfo {author} {\bibfnamefont {S.~C.}\ \bibnamefont {Glotzer}}, \ and\
  \bibinfo {author} {\bibfnamefont {D.~J.}\ \bibnamefont {Durian}},\
  }\href@noop {} {\bibfield  {journal} {\bibinfo  {journal} {Nature Phys.}\
  }\textbf {\bibinfo {volume} {3}},\ \bibinfo {pages} {260} (\bibinfo {year}
  {2007})}\BibitemShut {NoStop}%
\bibitem [{\citenamefont {Abate}\ and\ \citenamefont {Durian}(2007)}]{air2}%
  \BibitemOpen
  \bibfield  {author} {\bibinfo {author} {\bibfnamefont {A.~R.}\ \bibnamefont
  {Abate}}\ and\ \bibinfo {author} {\bibfnamefont {D.~J.}\ \bibnamefont
  {Durian}},\ }\href@noop {} {\bibfield  {journal} {\bibinfo  {journal} {Phys.\
  Rev.\ E}\ }\textbf {\bibinfo {volume} {76}},\ \bibinfo {pages} {021306}
  (\bibinfo {year} {2007})}\BibitemShut {NoStop}%
\bibitem [{\citenamefont {Lechenault}\ \emph {et~al.}(2008)\citenamefont
  {Lechenault}, \citenamefont {Dauchot}, \citenamefont {Biroli},\ and\
  \citenamefont {Bouchaud}}]{horiz1}%
  \BibitemOpen
  \bibfield  {author} {\bibinfo {author} {\bibfnamefont {F.}~\bibnamefont
  {Lechenault}}, \bibinfo {author} {\bibfnamefont {O.}~\bibnamefont {Dauchot}},
  \bibinfo {author} {\bibfnamefont {G.}~\bibnamefont {Biroli}}, \ and\ \bibinfo
  {author} {\bibfnamefont {J.~P.}\ \bibnamefont {Bouchaud}},\ }\href@noop {}
  {\bibfield  {journal} {\bibinfo  {journal} {Euro.\ Phys.\ Lett.}\ }\textbf
  {\bibinfo {volume} {83}},\ \bibinfo {pages} {46003} (\bibinfo {year}
  {2008})}\BibitemShut {NoStop}%
\bibitem [{\citenamefont {Lechenault}\ \emph {et~al.}(2010)\citenamefont
  {Lechenault}, \citenamefont {Candelier}, \citenamefont {Dauchot},
  \citenamefont {Bouchaud},\ and\ \citenamefont {Biroli}}]{horiz0}%
  \BibitemOpen
  \bibfield  {author} {\bibinfo {author} {\bibfnamefont {F.}~\bibnamefont
  {Lechenault}}, \bibinfo {author} {\bibfnamefont {R.}~\bibnamefont
  {Candelier}}, \bibinfo {author} {\bibfnamefont {O.}~\bibnamefont {Dauchot}},
  \bibinfo {author} {\bibfnamefont {J.~P.}\ \bibnamefont {Bouchaud}}, \ and\
  \bibinfo {author} {\bibfnamefont {G.}~\bibnamefont {Biroli}},\ }\href@noop {}
  {\bibfield  {journal} {\bibinfo  {journal} {Soft Matter}\ }\textbf {\bibinfo
  {volume} {6}},\ \bibinfo {pages} {3059} (\bibinfo {year} {2010})}\BibitemShut
  {NoStop}%
\bibitem [{\citenamefont {Mayer}\ \emph {et~al.}(2004)\citenamefont {Mayer},
  \citenamefont {Bissig}, \citenamefont {Berthier}, \citenamefont {Cipelletti},
  \citenamefont {Garrahan}, \citenamefont {Sollich},\ and\ \citenamefont
  {Trappe}}]{foamco}%
  \BibitemOpen
  \bibfield  {author} {\bibinfo {author} {\bibfnamefont {P.}~\bibnamefont
  {Mayer}}, \bibinfo {author} {\bibfnamefont {H.}~\bibnamefont {Bissig}},
  \bibinfo {author} {\bibfnamefont {L.}~\bibnamefont {Berthier}}, \bibinfo
  {author} {\bibfnamefont {L.}~\bibnamefont {Cipelletti}}, \bibinfo {author}
  {\bibfnamefont {J.~P.}\ \bibnamefont {Garrahan}}, \bibinfo {author}
  {\bibfnamefont {P.}~\bibnamefont {Sollich}}, \ and\ \bibinfo {author}
  {\bibfnamefont {V.}~\bibnamefont {Trappe}},\ }\href@noop {} {\bibfield
  {journal} {\bibinfo  {journal} {Phys.\ Rev.\ Lett.}\ }\textbf {\bibinfo
  {volume} {93}},\ \bibinfo {pages} {115701} (\bibinfo {year}
  {2004})}\BibitemShut {NoStop}%
\bibitem [{\citenamefont {Gluckman}\ \emph {et~al.}(1995)\citenamefont
  {Gluckman}, \citenamefont {Arnold},\ and\ \citenamefont
  {Gollub}}]{chaoticwaves}%
  \BibitemOpen
  \bibfield  {author} {\bibinfo {author} {\bibfnamefont {B.~J.}\ \bibnamefont
  {Gluckman}}, \bibinfo {author} {\bibfnamefont {C.~B.}\ \bibnamefont
  {Arnold}}, \ and\ \bibinfo {author} {\bibfnamefont {J.~P.}\ \bibnamefont
  {Gollub}},\ }\href@noop {} {\bibfield  {journal} {\bibinfo  {journal} {Phys.
  Rev. E}\ }\textbf {\bibinfo {volume} {51}},\ \bibinfo {pages} {1128}
  (\bibinfo {year} {1995})}\BibitemShut {NoStop}%
\bibitem [{\citenamefont {Vella}\ and\ \citenamefont
  {Mahadevan}(2005)}]{Cheerios}%
  \BibitemOpen
  \bibfield  {author} {\bibinfo {author} {\bibfnamefont {D.}~\bibnamefont
  {Vella}}\ and\ \bibinfo {author} {\bibfnamefont {L.}~\bibnamefont
  {Mahadevan}},\ }\href@noop {} {\bibfield  {journal} {\bibinfo  {journal} {Am.
  J. Phys.}\ }\textbf {\bibinfo {volume} {73}},\ \bibinfo {pages} {817}
  (\bibinfo {year} {2005})}\BibitemShut {NoStop}%
\bibitem [{\citenamefont {Chan}\ \emph {et~al.}(1981)\citenamefont {Chan},
  \citenamefont {J.~D.~Henry},\ and\ \citenamefont {White}}]{Capillaryforce}%
  \BibitemOpen
  \bibfield  {author} {\bibinfo {author} {\bibfnamefont {D.~Y.~C.}\
  \bibnamefont {Chan}}, \bibinfo {author} {\bibfnamefont {J.}~\bibnamefont
  {J.~D.~Henry}}, \ and\ \bibinfo {author} {\bibfnamefont {L.~R.}\ \bibnamefont
  {White}},\ }\href@noop {} {\bibfield  {journal} {\bibinfo  {journal} {J.
  Colloid Interface Sci.}\ }\textbf {\bibinfo {volume} {79}},\ \bibinfo {pages}
  {410} (\bibinfo {year} {1981})}\BibitemShut {NoStop}%
\bibitem [{\citenamefont {Herbst}\ \emph {et~al.}(2005)\citenamefont {Herbst},
  \citenamefont {Cafiero}, \citenamefont {Zippelius}, \citenamefont
  {Herrmann},\ and\ \citenamefont {Luding}}]{2d_driven_GM_simulation}%
  \BibitemOpen
  \bibfield  {author} {\bibinfo {author} {\bibfnamefont {O.}~\bibnamefont
  {Herbst}}, \bibinfo {author} {\bibfnamefont {R.}~\bibnamefont {Cafiero}},
  \bibinfo {author} {\bibfnamefont {A.}~\bibnamefont {Zippelius}}, \bibinfo
  {author} {\bibfnamefont {H.~J.}\ \bibnamefont {Herrmann}}, \ and\ \bibinfo
  {author} {\bibfnamefont {S.}~\bibnamefont {Luding}},\ }\href@noop {}
  {\bibfield  {journal} {\bibinfo  {journal} {Phys.\ of Fluids}\ }\textbf
  {\bibinfo {volume} {17}},\ \bibinfo {pages} {107102} (\bibinfo {year}
  {2005})}\BibitemShut {NoStop}%
\bibitem [{\citenamefont {Nordstrom}\ \emph {et~al.}(2011)\citenamefont
  {Nordstrom}, \citenamefont {Gollub},\ and\ \citenamefont {Durian}}]{shear2}%
  \BibitemOpen
  \bibfield  {author} {\bibinfo {author} {\bibfnamefont {K.~N.}\ \bibnamefont
  {Nordstrom}}, \bibinfo {author} {\bibfnamefont {J.~P.}\ \bibnamefont
  {Gollub}}, \ and\ \bibinfo {author} {\bibfnamefont {D.~J.}\ \bibnamefont
  {Durian}},\ }\href@noop {} {\bibfield  {journal} {\bibinfo  {journal} {Phys.\
  Rev.\ E}\ }\textbf {\bibinfo {volume} {84}},\ \bibinfo {pages} {021403}
  (\bibinfo {year} {2011})}\BibitemShut {NoStop}%
\bibitem [{\citenamefont {Berthier}(2004)}]{BerthierPRE2004}%
  \BibitemOpen
  \bibfield  {author} {\bibinfo {author} {\bibfnamefont {L.}~\bibnamefont
  {Berthier}},\ }\href@noop {} {\bibfield  {journal} {\bibinfo  {journal}
  {Phys. Rev. E (R)}\ }\textbf {\bibinfo {volume} {69}},\ \bibinfo {pages}
  {020201} (\bibinfo {year} {2004})}\BibitemShut {NoStop}%
\bibitem [{\citenamefont {Berthier}\ \emph {et~al.}(2005)\citenamefont
  {Berthier}, \citenamefont {Biroli}, \citenamefont {Bouchaud}, \citenamefont
  {Cipelletti}, \citenamefont {Masri}, \citenamefont {L'H\^{o}te},
  \citenamefont {Ladieu},\ and\ \citenamefont
  {Pierno}}]{BerthierBiroliBouchaudCipellettiMasriLHoteLadieuPierno}%
  \BibitemOpen
  \bibfield  {author} {\bibinfo {author} {\bibfnamefont {L.}~\bibnamefont
  {Berthier}}, \bibinfo {author} {\bibfnamefont {G.}~\bibnamefont {Biroli}},
  \bibinfo {author} {\bibfnamefont {J.-P.}\ \bibnamefont {Bouchaud}}, \bibinfo
  {author} {\bibfnamefont {L.}~\bibnamefont {Cipelletti}}, \bibinfo {author}
  {\bibfnamefont {D.~E.}\ \bibnamefont {Masri}}, \bibinfo {author}
  {\bibfnamefont {D.}~\bibnamefont {L'H\^{o}te}}, \bibinfo {author}
  {\bibfnamefont {F.}~\bibnamefont {Ladieu}}, \ and\ \bibinfo {author}
  {\bibfnamefont {M.}~\bibnamefont {Pierno}},\ }\href@noop {} {\bibfield
  {journal} {\bibinfo  {journal} {Science}\ }\textbf {\bibinfo {volume}
  {310}},\ \bibinfo {pages} {1797} (\bibinfo {year} {2005})}\BibitemShut
  {NoStop}%
\bibitem [{\citenamefont {Berthier}\ \emph {et~al.}(2011)\citenamefont
  {Berthier}, \citenamefont {Biroli}, \citenamefont {Bouchaud}, \citenamefont
  {Cipelletti},\ and\ \citenamefont {van
  Saarloos}}]{BerthierBiroliBouchaudCipellettiSaarloos}%
  \BibitemOpen
  \bibfield  {author} {\bibinfo {author} {\bibfnamefont {L.}~\bibnamefont
  {Berthier}}, \bibinfo {author} {\bibfnamefont {G.}~\bibnamefont {Biroli}},
  \bibinfo {author} {\bibfnamefont {J.~P.}\ \bibnamefont {Bouchaud}}, \bibinfo
  {author} {\bibfnamefont {L.}~\bibnamefont {Cipelletti}}, \ and\ \bibinfo
  {author} {\bibfnamefont {W.}~\bibnamefont {van Saarloos}},\ }\href@noop {}
  {\bibfield  {journal} {\bibinfo  {journal} {Dynamical Heterogeneities in
  Glasses, Colloids, and Granular Media (Oxford University Press, Oxford)}\ }
  (\bibinfo {year} {2011})}\BibitemShut {NoStop}%
\bibitem [{\citenamefont {Hatano}(2011)}]{Hatano_scaling_heterogeneity}%
  \BibitemOpen
  \bibfield  {author} {\bibinfo {author} {\bibfnamefont {T.}~\bibnamefont
  {Hatano}},\ }\href@noop {} {\bibfield  {journal} {\bibinfo  {journal} {J.
  Phys.: Conf. Ser.}\ }\textbf {\bibinfo {volume} {319}},\ \bibinfo {pages}
  {012011} (\bibinfo {year} {2011})}\BibitemShut {NoStop}%
\bibitem [{\citenamefont {Marty}\ and\ \citenamefont {Dauchot}(2005)}]{shear1}%
  \BibitemOpen
  \bibfield  {author} {\bibinfo {author} {\bibfnamefont {G.}~\bibnamefont
  {Marty}}\ and\ \bibinfo {author} {\bibfnamefont {O.}~\bibnamefont
  {Dauchot}},\ }\href@noop {} {\bibfield  {journal} {\bibinfo  {journal}
  {Phys.\ Rev.\ Lett.}\ }\textbf {\bibinfo {volume} {94}},\ \bibinfo {pages}
  {015701} (\bibinfo {year} {2005})}\BibitemShut {NoStop}%
\bibitem [{\citenamefont {Katsuragi}\ \emph {et~al.}(2010)\citenamefont
  {Katsuragi}, \citenamefont {Abate},\ and\ \citenamefont {Durian}}]{shear3}%
  \BibitemOpen
  \bibfield  {author} {\bibinfo {author} {\bibfnamefont {H.}~\bibnamefont
  {Katsuragi}}, \bibinfo {author} {\bibfnamefont {A.~R.}\ \bibnamefont
  {Abate}}, \ and\ \bibinfo {author} {\bibfnamefont {D.~J.}\ \bibnamefont
  {Durian}},\ }\href@noop {} {\bibfield  {journal} {\bibinfo  {journal} {Soft
  Matter}\ }\textbf {\bibinfo {volume} {6}},\ \bibinfo {pages} {3023} (\bibinfo
  {year} {2010})}\BibitemShut {NoStop}%
\bibitem [{\citenamefont {Abate}\ and\ \citenamefont {Durian}(2006)}]{air1}%
  \BibitemOpen
  \bibfield  {author} {\bibinfo {author} {\bibfnamefont {A.~R.}\ \bibnamefont
  {Abate}}\ and\ \bibinfo {author} {\bibfnamefont {D.~J.}\ \bibnamefont
  {Durian}},\ }\href@noop {} {\bibfield  {journal} {\bibinfo  {journal} {Phys.\
  Rev.\ E}\ }\textbf {\bibinfo {volume} {74}},\ \bibinfo {pages} {031308}
  (\bibinfo {year} {2006})}\BibitemShut {NoStop}%
\bibitem [{\citenamefont {Chaikin}\ and\ \citenamefont
  {Lubensky}(1995)}]{DC-refI}%
  \BibitemOpen
  \bibfield  {author} {\bibinfo {author} {\bibfnamefont {P.~M.}\ \bibnamefont
  {Chaikin}}\ and\ \bibinfo {author} {\bibfnamefont {T.~C.}\ \bibnamefont
  {Lubensky}},\ }\href@noop {} {\emph {\bibinfo {title} {Principles of
  Condensed Matter Physics}}}\ (\bibinfo  {publisher} {Cambridge University
  Press, Cambridge},\ \bibinfo {year} {1995})\BibitemShut {NoStop}%
\bibitem [{\citenamefont {Nishimori}\ and\ \citenamefont
  {Ortiz}(2011)}]{DC-refII}%
  \BibitemOpen
  \bibfield  {author} {\bibinfo {author} {\bibfnamefont {H.}~\bibnamefont
  {Nishimori}}\ and\ \bibinfo {author} {\bibfnamefont {G.}~\bibnamefont
  {Ortiz}},\ }\href@noop {} {\emph {\bibinfo {title} {Elements of Phase
  Transitions and Critical Phenomena}}}\ (\bibinfo  {publisher} {Oxford
  University Press, Oxford},\ \bibinfo {year} {2011})\BibitemShut {NoStop}%
\bibitem [{\citenamefont {Ikeda}\ \emph {et~al.}(2013)\citenamefont {Ikeda},
  \citenamefont {Berthier},\ and\ \citenamefont {Biroli}}]{DC_refIII}%
  \BibitemOpen
  \bibfield  {author} {\bibinfo {author} {\bibfnamefont {A.}~\bibnamefont
  {Ikeda}}, \bibinfo {author} {\bibfnamefont {L.}~\bibnamefont {Berthier}}, \
  and\ \bibinfo {author} {\bibfnamefont {G.}~\bibnamefont {Biroli}},\
  }\href@noop {} {\bibfield  {journal} {\bibinfo  {journal} {J. Chem. Phys.}\
  }\textbf {\bibinfo {volume} {138}},\ \bibinfo {pages} {12A507} (\bibinfo
  {year} {2013})}\BibitemShut {NoStop}%
\bibitem [{\citenamefont {Hatano}(2009)}]{Hatano_scaling_growingscales}%
  \BibitemOpen
  \bibfield  {author} {\bibinfo {author} {\bibfnamefont {T.}~\bibnamefont
  {Hatano}},\ }\href@noop {} {\bibfield  {journal} {\bibinfo  {journal} {Phys.
  Rev. E}\ }\textbf {\bibinfo {volume} {79}},\ \bibinfo {pages} {050301(R)}
  (\bibinfo {year} {2009})}\BibitemShut {NoStop}%
\bibitem [{\citenamefont {Tighe}\ \emph {et~al.}(2010)\citenamefont {Tighe},
  \citenamefont {Woldhuis}, \citenamefont {Remmers}, \citenamefont {van
  Saarloos},\ and\ \citenamefont {van Hecke}}]{Brian_scaling_fluctuations}%
  \BibitemOpen
  \bibfield  {author} {\bibinfo {author} {\bibfnamefont {B.~P.}\ \bibnamefont
  {Tighe}}, \bibinfo {author} {\bibfnamefont {E.}~\bibnamefont {Woldhuis}},
  \bibinfo {author} {\bibfnamefont {J.~J.~C.}\ \bibnamefont {Remmers}},
  \bibinfo {author} {\bibfnamefont {W.}~\bibnamefont {van Saarloos}}, \ and\
  \bibinfo {author} {\bibfnamefont {M.}~\bibnamefont {van Hecke}},\ }\href@noop
  {} {\bibfield  {journal} {\bibinfo  {journal} {Phys. Rev. Lett.}\ }\textbf
  {\bibinfo {volume} {105}},\ \bibinfo {pages} {088303} (\bibinfo {year}
  {2010})}\BibitemShut {NoStop}%
\bibitem [{\citenamefont {Hatano}(2010)}]{Hatano_criticalscaling}%
  \BibitemOpen
  \bibfield  {author} {\bibinfo {author} {\bibfnamefont {T.}~\bibnamefont
  {Hatano}},\ }\href@noop {} {\bibfield  {journal} {\bibinfo  {journal} {Prog.
  Theor. Phys. Supplement}\ }\textbf {\bibinfo {volume} {184}},\ \bibinfo
  {pages} {143} (\bibinfo {year} {2010})}\BibitemShut {NoStop}%
\bibitem [{\citenamefont {Douady}(1990)}]{Douady}%
  \BibitemOpen
  \bibfield  {author} {\bibinfo {author} {\bibfnamefont {S.}~\bibnamefont
  {Douady}},\ }\href@noop {} {\bibfield  {journal} {\bibinfo  {journal} {J.
  Fluid Mech.}\ }\textbf {\bibinfo {volume} {221}},\ \bibinfo {pages} {383}
  (\bibinfo {year} {1990})}\BibitemShut {NoStop}%
\bibitem [{Note1()}]{Note1}%
  \BibitemOpen
  \bibinfo {note} {The particles have been custom made by a collaborating
  company and are not commercially available.}\BibitemShut {Stop}%
\bibitem [{\citenamefont {Luding}(2001)}]{poly_luding}%
  \BibitemOpen
  \bibfield  {author} {\bibinfo {author} {\bibfnamefont {S.}~\bibnamefont
  {Luding}},\ }\href@noop {} {\bibfield  {journal} {\bibinfo  {journal} {Advs.\
  Complex Syst.}\ }\textbf {\bibinfo {volume} {04}},\ \bibinfo {pages} {379}
  (\bibinfo {year} {2001})}\BibitemShut {NoStop}%
\bibitem [{\citenamefont {Duez}\ \emph {et~al.}(2007)\citenamefont {Duez},
  \citenamefont {Ybert}, \citenamefont {Clanet},\ and\ \citenamefont
  {Bocquet}}]{cleaningprotocol}%
  \BibitemOpen
  \bibfield  {author} {\bibinfo {author} {\bibfnamefont {C.}~\bibnamefont
  {Duez}}, \bibinfo {author} {\bibfnamefont {C.}~\bibnamefont {Ybert}},
  \bibinfo {author} {\bibfnamefont {C.}~\bibnamefont {Clanet}}, \ and\ \bibinfo
  {author} {\bibfnamefont {L.}~\bibnamefont {Bocquet}},\ }\href@noop {}
  {\bibfield  {journal} {\bibinfo  {journal} {Nat. Phys.}\ }\textbf {\bibinfo
  {volume} {3}},\ \bibinfo {pages} {180} (\bibinfo {year} {2007})}\BibitemShut
  {NoStop}%
\bibitem [{Note2()}]{Note2}%
  \BibitemOpen
  \bibinfo {note} {Even in the dilute case, the distance $r$ between the
  floaters is much smaller than the capillary length $L_c=\protect \sqrt
  {\gamma /\rho _l g}$ with $\gamma $ the surface tension coefficient of the
  interface, $\rho _l$ the liquid density, and $g$ the acceleration of gravity.
  For an air-water interface at $20\protect \tmspace +\thinmuskip
  {.1667em}^\circ $C, $L_c=2.7$ mm.}\BibitemShut {Stop}%
\bibitem [{\citenamefont {Cicuta}\ and\ \citenamefont
  {Vella}(2009)}]{bucklingII}%
  \BibitemOpen
  \bibfield  {author} {\bibinfo {author} {\bibfnamefont {P.}~\bibnamefont
  {Cicuta}}\ and\ \bibinfo {author} {\bibfnamefont {D.}~\bibnamefont {Vella}},\
  }\href@noop {} {\bibfield  {journal} {\bibinfo  {journal} {Phys. Rev. Lett.}\
  }\textbf {\bibinfo {volume} {102}},\ \bibinfo {pages} {138302} (\bibinfo
  {year} {2009})}\BibitemShut {NoStop}%
\bibitem [{\citenamefont {Goldhirsch}(2010)}]{coarse6}%
  \BibitemOpen
  \bibfield  {author} {\bibinfo {author} {\bibfnamefont {I.}~\bibnamefont
  {Goldhirsch}},\ }\href@noop {} {\bibfield  {journal} {\bibinfo  {journal}
  {Granular Matter}\ }\textbf {\bibinfo {volume} {12}},\ \bibinfo {pages} {239}
  (\bibinfo {year} {2010})}\BibitemShut {NoStop}%
\bibitem [{\citenamefont {Weinhart}\ \emph {et~al.}(2012)\citenamefont
  {Weinhart}, \citenamefont {Thornton}, \citenamefont {Luding},\ and\
  \citenamefont {Bokhove}}]{coarse9}%
  \BibitemOpen
  \bibfield  {author} {\bibinfo {author} {\bibfnamefont {T.}~\bibnamefont
  {Weinhart}}, \bibinfo {author} {\bibfnamefont {A.~R.}\ \bibnamefont
  {Thornton}}, \bibinfo {author} {\bibfnamefont {S.}~\bibnamefont {Luding}}, \
  and\ \bibinfo {author} {\bibfnamefont {O.}~\bibnamefont {Bokhove}},\
  }\href@noop {} {\bibfield  {journal} {\bibinfo  {journal} {Granular Matter}\
  }\textbf {\bibinfo {volume} {14}},\ \bibinfo {pages} {289} (\bibinfo {year}
  {2012})}\BibitemShut {NoStop}%
\bibitem [{\citenamefont {Goldenberg}\ and\ \citenamefont
  {Goldhirsch}(2006)}]{coarse10}%
  \BibitemOpen
  \bibfield  {author} {\bibinfo {author} {\bibfnamefont {C.}~\bibnamefont
  {Goldenberg}}\ and\ \bibinfo {author} {\bibfnamefont {I.}~\bibnamefont
  {Goldhirsch}},\ }\href@noop {} {\emph {\bibinfo {title} {Handbook of
  Theoretical and Computational Nanotechnology}}}\ (\bibinfo  {publisher}
  {American Scientific Publishers, Stevenson Ranch, CA},\ \bibinfo {year}
  {2006})\BibitemShut {NoStop}%
\bibitem [{Note3()}]{Note3}%
  \BibitemOpen
  \bibinfo {note} {The procedure in the ensemble average is to calculate
  $|\protect \mathbf {r}_i(t+\tau )-\protect \mathbf {r}_i(t)|$ using arbitrary
  starting times $t$ and averaging over t. A similar procedure is followed in
  calculating ensemble averages in the self-overlap order parameter, dynamic
  susceptibility and four-point correlation function.}\BibitemShut {Stop}%
\bibitem [{Note4()}]{Note4}%
  \BibitemOpen
  \bibinfo {note} {The optimal values for $d$ were obtained as follows: When
  looking at the subtraction procedure as a function of $d$ we find that the
  displacement rises steeply from zero for $d\ll \sigma $, $\sigma $ is the
  floater diameter, into a plateau from which it continues to rise. A value in
  the center of the plateau is chosen, which happens to correspond roughly to
  the particle diameter.}\BibitemShut {Stop}%
\bibitem [{Note5()}]{Note5}%
  \BibitemOpen
  \bibinfo {note} {From the (two-dimensional) pair correlation function $g(r)$
  we observe no evidence for significant crystallization which may cause an
  increase of this upper limit for $\phi _J$. Secondly, although buckling may
  be a significant factor, it does not lead to a broadening of the first peak
  in $g(r)$ that one would expect to be present if particles start to overlap
  for increasing $\phi $. And, finally, the homogenized local packing fraction
  shows a sharp cut-off at $\phi \approx 0.84$. These facts together suggests
  that $\phi _J<0.84$.}\BibitemShut {Stop}%
\bibitem [{Note6()}]{Note6}%
  \BibitemOpen
  \bibinfo {note} {Due to their limited range other functional forms could
  possibly also fit our dynamic time and length scales. However, we restrict
  ourselves to power-law fits in order to compare our results to those in the
  literature.}\BibitemShut {Stop}%
\bibitem [{\citenamefont {Berhanu}\ and\ \citenamefont
  {Kudrolli}(2010)}]{bucklingI}%
  \BibitemOpen
  \bibfield  {author} {\bibinfo {author} {\bibfnamefont {M.}~\bibnamefont
  {Berhanu}}\ and\ \bibinfo {author} {\bibfnamefont {A.}~\bibnamefont
  {Kudrolli}},\ }\href@noop {} {\bibfield  {journal} {\bibinfo  {journal}
  {Phys. Rev. Lett.}\ }\textbf {\bibinfo {volume} {105}},\ \bibinfo {pages}
  {098002} (\bibinfo {year} {2010})}\BibitemShut {NoStop}%
\bibitem [{Note7()}]{Note7}%
  \BibitemOpen
  \bibinfo {note} {Both $d$ and $a$ are first calculated for each $\phi $
  separately, and then appropriate values are obtained by averaging over the
  determined values.}\BibitemShut {Stop}%
\bibitem [{\citenamefont {La{\v c}evi{\' c}}\ \emph {et~al.}(2003)\citenamefont
  {La{\v c}evi{\' c}}, \citenamefont {Starr}, \citenamefont {Schr{\o}der},\
  and\ \citenamefont {Glotzer}}]{tech0}%
  \BibitemOpen
  \bibfield  {author} {\bibinfo {author} {\bibfnamefont {N.}~\bibnamefont
  {La{\v c}evi{\' c}}}, \bibinfo {author} {\bibfnamefont {F.~W.}\ \bibnamefont
  {Starr}}, \bibinfo {author} {\bibfnamefont {T.~B.}\ \bibnamefont
  {Schr{\o}der}}, \ and\ \bibinfo {author} {\bibfnamefont {S.~C.}\ \bibnamefont
  {Glotzer}},\ }\href@noop {} {\bibfield  {journal} {\bibinfo  {journal} {J.\
  Chem.\ Phys.}\ }\textbf {\bibinfo {volume} {119}},\ \bibinfo {pages} {7372}
  (\bibinfo {year} {2003})}\BibitemShut {NoStop}%
\end{thebibliography}%
\end{document}